\documentclass[sigplan, nonacm]{acmart}
\AtBeginDocument{%
  }

\usepackage[T1]{fontenc}
\usepackage[utf8]{inputenc}
\usepackage{amsmath,amssymb,bm}
\usepackage{comment}
\usepackage{booktabs,multirow}
\usepackage{threeparttable}
\usepackage{graphicx}
\usepackage{xspace}
\usepackage{url}
\usepackage{enumitem}
\usepackage{balance}
\usepackage{siunitx}
\usepackage{subcaption}
\usepackage{algorithm}
\usepackage{algpseudocode}
\usepackage{threeparttable}
\usepackage{textgreek}
\usepackage{multirow}
\usepackage[table]{xcolor}   
\usepackage{booktabs}        

\begin{document}

\newcommand{\System}{This work\xspace}
\newcommand{\system}{this work\xspace}
\newcommand{\TensorMD}{TensorMD\xspace}
\newcommand{\DPSE}{DP\xspace}
\newcommand{\DPA}{DPA-1\xspace}
\newcommand{\HEA}{$\mathsf{HEA}$\xspace}
\newcommand{\SSE}{$\mathsf{SSE}$\xspace}
\newcommand{\Sn}{$\mathsf{Sn}$\xspace}
\newcommand{\MoNi}{$\mathsf{MoNi}$\xspace}
\newcommand{\AlCuMg}{$\mathsf{AlCuMg}$\xspace}
\newcommand{\TODO}[1]{\textbf{[TODO: #1]}}

\title[]{Unlocking Multi-Component Bulk-Materials Molecular Dynamics   with a Small-Footprint Machine Learning Interatomic Potential}


\author{Yucheng Ouyang}
\authornote{Both authors contributed equally to this research.}
\orcid{0000-0001-8315-3667}
\affiliation{
  \institution{SKLP, ICT, CAS}
  \department{UCAS}
  \city{Beijing}
  \country{China}
}
\email{ouyangyucheng22z@ict.ac.cn}

\author{Xin Chen}
\authornotemark[1]
\authornote{Corresponding authors.}
\orcid{0000-0001-9643-0870}
\affiliation{
  \institution{National Key Laboratory of Computational Physics, Institute of Applied Physics and Computational Mathematics}
  \city{Beijing}
  \country{China}
}
\email{chen\_xin@iapcm.ac.cn}

\author{Ying Liu}
\authornotemark[2]
\orcid{0000-0003-1599-9665}
\affiliation{
  \institution{SKLP, ICT, CAS}
  \department{UCAS}
  \city{Beijing}
  \country{China}
}
\email{liuying2007@ict.ac.cn}

\author{Lifang Wang}
\affiliation{
  \institution{National Key Laboratory of Computational Physics, Institute of Applied Physics and Computational Mathematics}
  \city{Beijing}
  \country{China}
}
\email{wang\_lifang@iapcm.ac.cn}

\author{Xingyu Gao}
\affiliation{
  \institution{National Key Laboratory of Computational Physics, Institute of Applied Physics and Computational Mathematics}
  \city{Beijing}
  \country{China}
}
\email{gao\_xingyu@iapcm.ac.cn}

\author{Xiawei Du}
\affiliation{
  \institution{Tsinghua University}
  \department{Department of Computer Science and Technology}
  \city{Beijing}
  \country{China}
}
\email{duxw25@mails.tsinghua.edu.cn}

\author{Jianierken Habudelihan}
\affiliation{
  \institution{SKLP, ICT, CAS}
  \department{UCAS}
  \city{Beijing}
  \country{China}
}
\email{jianierken23s@ict.ac.cn}

\author{Haifeng Song}
\orcid{0000-0001-6215-0125}
\affiliation{
  \institution{National Key Laboratory of Computational Physics, Institute of Applied Physics and Computational Mathematics}
  \city{Beijing}
  \country{China}
}
\email{song\_haifeng@iapcm.ac.cn}

\author{Huimin Cui}
\affiliation{
  \institution{SKLP, ICT, CAS}
  \department{UCAS}
  \city{Beijing}
  \country{China}
}
\email{cuihm@ict.ac.cn}

\author{Xiaobing Feng}
\affiliation{
  \institution{SKLP, ICT, CAS}
  \department{UCAS}
  \city{Beijing}
  \country{China}
}
\email{fxb@ict.ac.cn}

\author{Jingling Xue}
\affiliation{
  \institution{University of New South Wales}
  \city{Sydney}
  \country{Australia}
}
\email{j.xue@unsw.edu.au}

\renewcommand{\shortauthors}{Yucheng Ouyang, Xin Chen, et al.}


\begin{abstract}

 Bulk materials, as opposed to nanomaterials, require molecular dynamics (MD) simulations on a large spatial scale ($\sim 10^9$ atoms or more) to adequately capture their atomic-scale physical properties. Previously, the introduction of machine-learning interatomic potentials (MLIPs) has extended MD to this scale, but even single-component bulk systems require tens of thousands of GPUs on high-end supercomputers. However, multi-component bulk MD simulations remain barely achievable, as the HBM footprint of existing MLIPs — already substantial for single-component systems — grows explosively in multi-component scenarios.

 This paper proposes an MLIP with a small HBM footprint  -- less than 3\% that of existing MLIPs -- unlocking multi-component bulk MD using only hundreds of GPUs.
 This is achieved by first identifying feature vectors and intermediate tensors as the two primary contributors to  HBM footprints in existing MLIPs. To address these two sources, 
 the dimensionality of the feature vectors has been reduced by introducing physical and chemical knowledge, and {intermediate tensors have been eliminated by aggressively fusing all kernels into a single mega-kernel.}

 In evaluation, the proposed MLIP has used 144 NVIDIA A100 GPUs to perform MD simulations on a  6-component bulk system with {$1.14\times 10^9$} atoms, while previously such MD simulation spatial scale  has been restricted to unary systems and typically achieved on high-end supercomputers equipped with tens of thousands of GPUs. 

\end{abstract}

\maketitle

\section{Introduction}
\label{sec:intro}


Bulk materials, as opposed to nanomaterials, are material volumes sufficiently large that their response is governed primarily by interior microstructure rather than by surfaces or nanoscale size effects. In this regime, quantum confinement effects are insignificant, and physical properties are size-independent. Consequently, bulk-material simulations are indispensable across various fields, such as the development of radiation-resistant alloys for nuclear energy~\cite{nuclear_ref} and the investigation of macroscopic mechanical responses under extreme conditions~\cite{DP_Sn}.
As in many other areas of physics research,  Molecular Dynamics (MD) has been a primary tool for studying the response of condensed matter in bulk-material simulation.

In practice, the {scale of MD simulation -- especially spatial scale -- is not a luxury but a necessity}, since larger dimensions are required to minimize surface effects and statistical fluctuations.
In particular, billion-atom ($\sim 10^9$) MD simulations are critical to make direct connection with experiment~\cite{snap_gb_sc21}.
For example in bulk metallic and alloy systems,
the simulated spatial scale is preferred to reach 0.1-1 $\mu m$  to effectively resolve defect kinetics and wave propagation, so that the simulated results can be practically used in shock loading~\cite{md_application_shock},
plasticity~\cite{md_application_shear_strength,md_application_hardening}, radiation damage~\cite{nuclear_ref}, dislocation~\cite{md_application_dislocation}, additive manufacturing\cite{md_application_AM} and multiscale coupling to continuum mechanics~\cite{md_application_deform}.

\begin{table}
\centering
\small
\setlength{\tabcolsep}{4pt}
\caption{State-of-the-art MD simulations of bulk materials: the MD spatial scale is determined by the per-atom HBM footprint of the MLIP used.}
\label{tbl:intro}
\begin{threeparttable}
\begin{tabular}{c|ccc}
\toprule
\multirow{2}{*}{\textsc{MLIPs}} & {\textsc{\# of atom}} & {\textsc{Per-atom}}  & \textsc{MD spatial} \\
& \textsc{species} & \textsc{ footprint} & \textsc{scale} \\
\midrule
\multirow{2}{*}{DP~\cite{DP_TC}} & \multirow{2}{*}{1} & {696.9 KB\tnote{a}}  & $1.3\times 10^{10}$ \\
& & \multicolumn{2}{c}{\color{gray}(on 90,000 accelerators)} \\
\midrule
\multirow{2}{*}{DP~\cite{DP_Ppopp}} & \multirow{2}{*}{1 } & 135.0 KB  & $3.4\times 10^9$ \\
& & \multicolumn{2}{c}{{\color{gray}(on 27,360 GPUs)}} \\
\midrule
\multirow{3}{*}{\TensorMD~\cite{TensorMD}} & \multirow{2}{*}{1 } & {67.1 KB}  & $4.0\times 10^9$ \\
& & \multicolumn{2}{c}{\color{gray}(on 16,000 GPUs)} \\
& {6 } & 365.6 KB\tnote{b}  & $1.5\times 10^{7}$ \\
\midrule
\DPSE~\cite{DPkit-v3,DP_Sn} & {1 } & {38.1 KB\tnote{b}}  & $1.0\times 10^{8}$  \\
(DeePMD-kit v3.1.2)& {6 } & 204.6 KB\tnote{b} & $2.6\times 10^{7}$  \\
\midrule
\multirow{3}{*}{\textbf{This work}}  & {1 } & \textbf{3.7 KB}  & $1.6\times 10^{9}$ \\
 & {6 } & \textbf{5.1 KB}  & $1.1\times 10^{9}$ \\
& & \multicolumn{2}{c}{\color{gray}(each on 144 GPUs)} \\
\bottomrule
\end{tabular}
\begin{tablenotes}[flushleft]
\footnotesize
\item[a] Results of this MLIP are reported for mixed precision of FP32 and FP64. Results of all other MLIPs are reported for FP64.
\item[b] Measured in this work with 144 GPUs.
\end{tablenotes}
\end{threeparttable}
\end{table}



MD simulations necessitate both accurate and computationally efficient predictions of potential energy and atomic forces to faithfully describe interactions within complex systems~\cite{Allegro}. Traditionally, first-principles methods offer high accuracy but are computationally expensive. With the integration of AI and HPC, modern machine learning interatomic potentials (MLIPs)\cite{DP_GB,BPNN,snap_hdc_prb,Allegro} have been developed, significantly expanding the accessible system sizes. 


However, as shown in Table~\ref{tbl:intro}, 
SOTA MLIP MD simulations of bulk materials are confined to single-component  systems and are quite expensive.
For example, DP~\cite{DP_TC} has simulated $1.3\times 10^{10}$ $\mathsf{Cu}$ atoms, consuming $84\%$ of the new generation Sunway supercomputer ($90,000$ many-core accelerators); 
\TensorMD~\cite{TensorMD} has achieved  $4\times 10^{9}$-$\mathsf{W}$-atom MD simulation, occupying two-thirds of ORISE supercomputer ($16,000$ GPUs). 
The reason behind this is the \textit{huge per-atom HBM footprint} of existing MLIPs, 
which limits the number of atoms that can be processed on each GPU.
Moreover, for multi-component systems, the footprint  grows explosively -- DP~\cite{DPkit-v3,DP_Sn} requires 38.1 KB/atom for  the single-component $\mathsf{Sn}$ system but 204.6 KB/atom for the 6-component high-entropy alloy -- making  multi-component bulk-materials MD simulation barely feasible even on  leading supercomputers.


The reason behind such memory consumption is rooted in the nature of MD, in which the dynamics of each atom is determined by its neighboring atoms.
Specifically, for each atom, MLIPs must keep track of all its neighbors (typically hundreds for bulk materials due to their high density), abstracting the interaction between the central atom and each neighbor into a feature vector to describe their pairwise interactions. This results in a large per-atom HBM footprint, {especially for multi-component systems, in which feature vectors need to be populated to differentiate the atom species of neighboring atoms}. 
Furthermore, MLIPs necessitate that model inference — calculating potential energy from atomic coordinates — be immediately followed by auto-differentiation — computing atomic forces as the gradients of the potential energy with respect to coordinates, referred to as the forward and backward phases, respectively. This requires a large number of intermediate tensors to be retained across the two phases, further exacerbating the already substantial per-atom HBM footprint.
Several prior works have attempted to reduce this per-atom footprint through dynamic memory allocation of neighboring atoms~\cite{TensorMD} and aggressive kernel fusion~\cite{DP_Ppopp,TensorMD}; however as shown in Table~\ref{tbl:intro}, the results remained modest before this work.



In this paper, we have proposed an MLIP with a significantly reduced per-atom memory footprint, amounting to only 3\% of that required by SOTA works for multi-component systems, as shown in Table~\ref{tbl:intro}. This small-footprint MLIP enables multi-component MD simulations  at the $\sim 10^9$ atom scale to be completed using only 144 GPUs, while previously single-component simulations are only feasible  on high-end supercomputers equipped with $\sim 20,000$ GPUs. Consequently, it makes practical bulk material research feasible on commonly accessible clusters.
This has been achieved by introducing  innovations  to address the 
the two primary contributors to HBM footprint, i.e., feature vectors and intermediate tensors, respectively.

This paper makes the following contributions:

\begin{itemize}
    \item \textbf{Reducing feature vector footprint}. 
    The two types of feature vectors -- specifically, the radial and angular feature vectors used to characterize the distances and directions of neighboring atoms for a given atom respectively -- are both compressed by leveraging physical knowledge to reduce the per-atom HBM footprint.  Dimensionality of the radial feature vector is reduced by distinguishing atomic species within multi-component systems, while dimensionality of the angular feature vector is reduced by eliminating the {dependencies among pair-wise atomic displacements.}

    \item \textbf{Eliminating intermediate tensor footprint}. 
    All calculations from both forward and backward phases have been encapsulated into one single mega-kernel, thereby achieving  minimal HBM footprint for MLIPs.
    This has been done by collaboratively applying tailored tiling, long-live-range variable re-computation, and warp-level data shuffle.

\item \textbf{Bulk-material evaluation}.
{The proposed MLIP has simulated  a 6-component of $1.14\times 10^9$ \HEA  atoms 
on 144 NVIDIA A100 GPUs,  while such MD simulation spatial scale  has been restricted to unary systems and typically achieved on high-end supercomputers equipped with tens of thousands of GPUs. }
{Compared with two SOTA MLIPs, i.e., \DPSE~\cite{DP_GB,DP_Ppopp,DPkit-v3} and \TensorMD~\cite{TensorMD}, the per-atom HBM footprint can be reduced to 2.50\% and 1.39\% of them respectively for \HEA, and the MD simulation throughput can be improved by 13.9$\times$ and 9.7$\times$  respectively.}

    
\end{itemize}




\section{Background}
\label{subsec:v1overview}
\subsection{Overview of MLIPs}




  \subsubsection{MLIP Training}

  {As shown in Fig.~\ref{fig:flowchar}(a)}, MLIPs are trained with datasets  calculated by the highly accurate first-principle methods such as ~\cite{DFT1,DFT2}, so that MLIPs are able to align with  their \textit{ab initio} accuracies. In particular, the datasets are composed of three parts of atomic coordinates, potential energies and atomic forces, 

 \subsubsection{MLIP Inference}



\begin{figure}
  \begin{center}
  \centering
    \begin{subfigure}[b]{0.48\textwidth}
        \centering
        \includegraphics[width=\columnwidth]{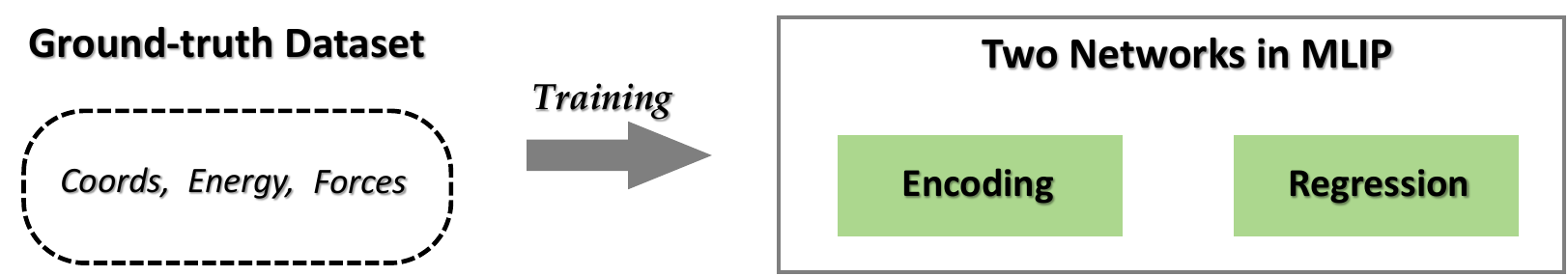}
        \Description{}
        \caption{MLIP training.}
        \label{fig:flowchart(a)}
    \end{subfigure}
    \hfill
    \begin{subfigure}[b]{0.48\textwidth}
        \centering
        \includegraphics[width=\linewidth]{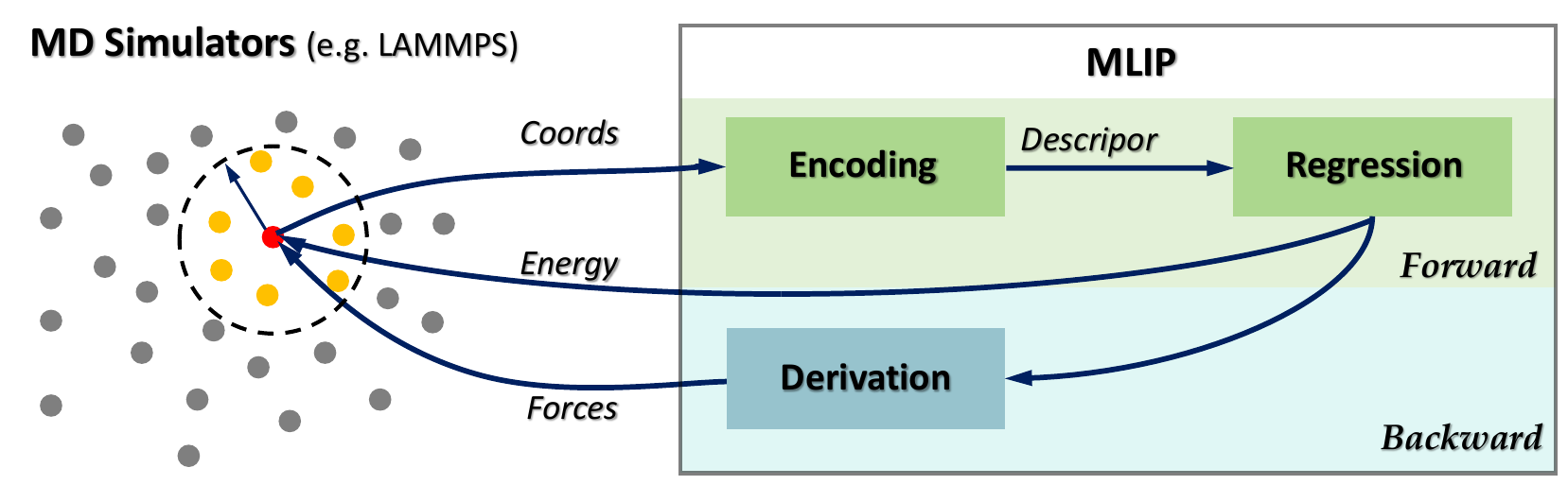}
        \caption{MLIP inference.}
        \label{fig:flowchart(b)}
    \end{subfigure}
    \caption{General workflow of MLIPs.}
    \label{fig:flowchar}
  \end{center}
\end{figure}

{As shown in Fig.~\ref{fig:flowchar}(b)}, during inference, MLIPs typically interface with MD simulators such as LAMMPS~\cite{LAMMPS} and GPUMD~\cite{GPUMD}. This interfacing occurs at each time step (typically $0.001$ picoseconds), during which the MLIP receives atomic coordinates from the MD simulator, computes potential energies and atomic forces, and then returns them to the MD simulator to update the atomic coordinates for the next time step.
MLIPs involve two phases of forward propagation and backward propagation, with the former computing potential energies from the atomic coordinates and the latter yielding atomic forces from the calculated potential energies.
The \text{forward propagation} can be further divided into two stages, with an $\mathsf{Encoding}$ stage constructing a descriptor for each atom from  the coordinates of all its neighboring atoms, and a $\mathsf{Regression}$ stage calculating the atom's potential energy from its descriptor. The backward propagation uses a $\mathsf{Derivation}$ stage to further compute the force of the atom by differentiating its potential energy with respect to its atomic coordinates.

\subsection{Analysis on MD Spatial Scales}




As stressed earlier, huge per-atom HBM footprint is the major obstacle preventing MD simulations from reaching the large spatial scales required by practical bulk-material research, and it has two major sources, i.e., the feature vectors that abstract the pairwise interatomic interactions  and the intermediate variables kept for differentiating energies to compute forces.

Table~\ref{tbl:breakdown} lists the per-atom HBM footprint breakdown for \TensorMD $\mathsf{HEA}$ simulation\cite{TensorMD}.
The reason to choose \TensorMD for such quantitative analysis is that, it has achieved SOTA performance and accuracy~\cite{TensorMD} by embedding rich physics knowledge~\cite{DP_Descriptor}~\cite{Baskes_MEAM_1989}~\cite{GEAM}  in its feature vector design, which is also used (and further optimized) in this work.



From the table, we can see that the $\mathsf{Encoding}$ and $\mathsf{Derivation}$ stages are most memory intensive.
In $\mathsf{Encoding}$, which constructs a descriptor for each atom from the coordinates of all its neighboring atoms, MLIPs typically keep track of each neighbor with one or more feature vectors, thereby there will be hundreds of vectors residing in GPU memory, with each vector typically including dozens of elements.
In $\mathsf{Derivation}$, which computes the force of an atom by differentiating its potential energy with respect to its atomic coordinates, large amount of intermediate tensors are required to be kept to compute gradients. 

\begin{table}[t]
\footnotesize
\centering
\setlength{\tabcolsep}{3pt}
\caption{Per-atom HBM footprint breakdown of performing MD simulation with \TensorMD on the \HEA system.}
\label{tbl:breakdown}
\resizebox{\columnwidth}{!}{%
\begin{tabular}{c|l|c}
\toprule
                       & Category                                          & \% of Total \\
\midrule
\multirow{3}{*}{MLIP}  & Feature vectors (in $\mathsf{Encoding}$)          & $42.8\%$   \\
                       & Intermediate tensors (for $\mathsf{Derivation}$)  & $52.2\%$   \\
                       & Others                                            & $4.1\%$    \\
\midrule
MD Simulator           & All                                               & $0.9\%$    \\
\bottomrule
\end{tabular}%
}
\end{table}

\section{Methodology and Innovations}
\label{sec:dim_red}

 This section proposes several innovations to reduce the two major memory consumers, i.e., feature vectors and intermediate tensors, together accounting for $\sim 95\%$ of MLIP footprint, as given in Table~\ref{tbl:breakdown}.

 After introducing the MLIP inference pipeline in Section~\ref{sec:Inference_Pipeline}, 
 strategies of reducing the per-atom HBM footprint consumed by feature vectors (more than $40\%$ in Table~\ref{tbl:breakdown}) are proposed in
 Sections~\ref{sec:Radial} and~\ref{sec:Angular}, with radial and angular feature vectors  reduced respectively. 
 Furthermore, approaches of reducing the per-atom HBM footprint consumed by intermediate tensors (more than $50\%$ in Table~\ref{tbl:breakdown}) are described in 
Section~\ref{sec:Intermediate}.
Finally, Section~\ref{sec:impl} summarizes implementation details in this work.


\begin{figure*}
  \centering
  \begin{subfigure}{0.69\textwidth}
    \centering
    \includegraphics[width=\textwidth]{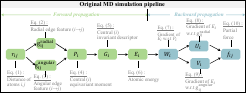}
    \caption{The MLIP inference pipeline of $\mathsf{Encoding}$-$\mathsf{Regression}$-$\mathsf{Derivation}$.}  
  \end{subfigure}\hfill
  \begin{subfigure}{0.29\textwidth}
    \centering
    \includegraphics[width=\textwidth]{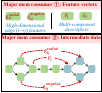}
    \caption{Major sources of HBM footprint.}  
  \end{subfigure}
\caption{The MLIP inference pipeline and its major sources of HBM footprint.}
\label{fig:fuse}
\end{figure*}

\begin{figure*}[t]
  \centering
  \begin{subfigure}{0.43\textwidth}
    \centering
    \includegraphics[width=\textwidth]{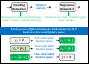}
    \caption{The encoding and regression networks can carry different species information. Introducing species info earlier in the  encoder exposes a lower-rank radial feature space.}  
  \end{subfigure}\hfill
  \begin{subfigure}{0.55\textwidth}
    \centering
    \includegraphics[width=\textwidth]{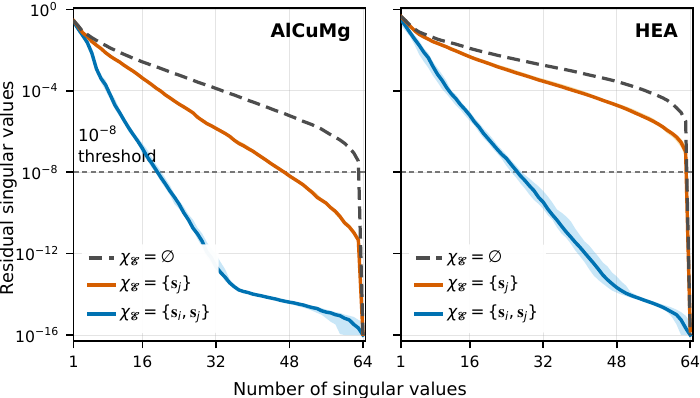}
    \caption{Singular-value distributions of radial feature matrices generated by encoders with different species-conditioning for $\mathsf{AlCuMg}$ and \HEA. Each matrix contains 50,001 rows, with each row being a 64-dimensional radial feature.}  
  \end{subfigure}
  \caption{Radial feature vector dimensionality reduction by early atom species discrimination.}
  \label{fig:svd_rank}
\end{figure*}

\subsection{Inference Pipeline}
\label{sec:Inference_Pipeline}


Figure~\ref{fig:fuse}(a) illustrates the inference pipeline of $\mathsf{Encoding}$-$\mathsf{Regression}$-$\mathsf{Derivation}$ used in this work, which is derived and improved from \TensorMD~\cite{TensorMD}. It is described as follows.


\subsubsection{The Encoding Stage.}

Given an atom $i$, let $\mathcal N(i)$ denote the set of its neighboring atoms   within a given cutoff.
Assume atom $j\in \mathcal N(i)$, given the position vectors  $\mathbf r_i, \mathbf r_j$ of atoms $i,j$,  { their interatomic distance $r_{ij}$ and unit distance vector $\hat{\mathbf r}_{ij}$ can be defined as} 
\begin{equation}
  \mathbf r_{ij}=\mathbf r_j-\mathbf r_i,\qquad
  r_{ij}=\lVert\mathbf r_{ij}\rVert,\qquad
  \hat{\mathbf r}_{ij}=\mathbf r_{ij}/r_{ij}.
  \label{eq:setup}
\end{equation}
Two feature vectors, i.e., the radial feature $\mathbf g^{\mathsf{radial}}_{ij}$ and the angular feature $\mathbf g^{\mathsf{angular}}_{ij}$, are constructed from  their interatomic distance $r_{ij}$ and unit distance vector $\hat{\mathbf r}_{ij}$ respectively, to characterize the distances and directions of atom $i$'s neighbors. In particular, the radial feature vector is calculated as
\begin{equation}
  \mathbf g^{\mathsf{radial}}_{ij}
  =\mathcal E\!\left(r_{ij};\chi_{\mathcal E}\right)
  \in\mathbb R^k
  \label{eq:radial}
\end{equation}
where $k$ is the radial feature dimensionality,  $\mathcal{E}$ is the trained encoding model, and $\chi_\mathcal E$ denotes the atom species information carried by model $\mathcal{E}$.

Under a maximum angular order $\ell_{max}\ge 0$, 
the angular feature vector $\mathbf g^{\mathsf{angular}}_{ij}$ can be written as 
\begin{equation}
  \mathbf g^{\mathsf{angular}}_{ij}
  =\Omega_{\ell_{max}}(\hat{\mathbf r}_{ij})\in\mathbb R^d,
  \label{eq:angular}
\end{equation}
where $d$ is the internal angular-state dimension.

Stacking the edge features as
$\mathbf G_{\mathsf{radial},i}\in\mathbb R^{|\mathcal N(i)|\times k}$ and
$\mathbf G_{\mathsf{angular},i}\in\mathbb R^{|\mathcal N(i)|\times d}$ gives the central equivariant moment
\begin{equation}
  \mathbf P_i
  =\mathbf G_{\mathsf{angular},i}^{\mathsf T}
   \mathbf G_{\mathsf{radial},i}
  =\sum_{j\in\mathcal N(i)}
   {{\mathbf g^{\mathsf{angular}}_{ij}}^{\mathsf T} \otimes {\mathbf g^{\mathsf{radial}}_{ij}}}
  \in\mathbb R^{d\times k}.
  \label{eq:equivariant}
\end{equation}
Relative coordinates make $\mathbf P_i$ translation invariant, the neighbor sum makes it invariant to neighbor ordering, and its angular blocks transform equivariantly under rotation. Partition the rows of $\mathbf P_i$ by angular order, and let $\mathbf p^{(\ell)}_{i,k}$ denote the order-$\ell$ block of its $k$-th column. The invariant descriptor keeps the scalar block and contracts every higher-order block to its squared norm,
\begin{equation}
  G_{i,\ell k}=
  \begin{cases}
    p^{(0)}_{i,k}, & \ell=0,\\
    {\mathbf p^{(\ell)}_{i,k}}^{\mathsf T}
    C_\ell\mathbf p^{(\ell)}_{i,k}, & 1\leq\ell\leq \ell_{max},
  \end{cases}
  \qquad \mathbf G_i\in\mathbb R^{(\ell_{max}+1)\times k}.
  \label{eq:descriptor}
\end{equation}
Here $C_\ell$ is the fixed metric of the angular basis: it contains the
Cartesian multiplicities for the canonical representation and is the identity
for the orthonormal irreducible representation.

\subsubsection{The Regression Stage}

Given the descriptor $\mathbf G_i$ of atom $i$, its potential energy $E_i$ is predicted as
\begin{equation}
    E_i=\mathcal R\!\left(\mathbf G_i;\chi_\mathcal R\right)
    \label{eq:theory_E}
\end{equation}
where $\mathcal{R}$ is the trained regression model, and $\chi_\mathcal R$ denotes  the atom species information carried by model $\mathcal{R}$.

\subsubsection{The Derivation Stage.}

The explicit backward path first propagates the energy gradient from the
invariant descriptor to the central moment state, and then to the radial and
angular edge features:
\begin{align}
  \mathbf W_i
    &\equiv \frac{\partial E_i}{\partial\mathbf P_i},
  & \\
  \mathbf U_i
    &\equiv \frac{\partial E_i}
      {\partial\mathbf G_{\mathsf{radial},i}}
     =\mathbf G_{\mathsf{angular},i}\mathbf W_i,\\[-1pt]
  \mathbf V_i
    &\equiv \frac{\partial E_i}
      {\partial\mathbf G_{\mathsf{angular},i}}
     =\mathbf G_{\mathsf{radial},i}\mathbf W_i^{\mathsf T}.
\label{eq:backward_wuv}
\end{align}

With $J^{\mathsf{radial}}_{ij}=\partial\mathbf g^{\mathsf{radial}}_{ij}/ \partial\mathbf r_{ij}$ and $J^{\mathsf{angular}}_{ij}=\partial\mathbf g^{\mathsf{angular}}_{ij}/ \partial\mathbf r_{ij}$, the partial forces are:
\begin{equation}
  f_{ij}=\frac{\partial E_i}{\partial\mathbf r_{ij}}
  ={J^{\mathsf{radial}}_{ij}}^{\mathsf T}\mathbf u_{ij}
  +{J^{\mathsf{angular}}_{ij}}^{\mathsf T}\mathbf v_{ij},
  \label{eq:force}
\end{equation}
Atomic forces are the scatter-sum of the partial forces.

\subsubsection{Major Sources of HBM Footprint.}

The formulation above exposes two major classes of live state, illustrated in
Figure~\ref{fig:fuse}(b).

\noindent\texttt{Edge-related state.}
An unfused implementation may materialize
$\mathbf g^{\mathsf{radial}}_{ij}$,
$\mathbf g^{\mathsf{angular}}_{ij}$, their adjoints $\mathbf U_i$ and
$\mathbf V_i$, and geometric derivatives used by Equation~\ref{eq:force}.
These arrays contain one row per neighbor and therefore scale as
$\mathcal O(|\mathcal N(i)|(k+d))$ per center atom. Their footprint is controlled
primarily by the $k$, $d$, and neighbor count.

\noindent\texttt{Atom-relate intermediate state.}
The forward and backward stages also share the $\mathbf P_i\in\mathbb R^{d\times k}$, its adjoint $\mathbf W_i\in\mathbb R^{d\times k}$, the invariant descriptor $\mathbf G_i\in\mathbb R^{(\ell_{max}+1)\times k}$, and neural-network activations. These states must either remain live across stages or be recomputed.

\subsection{Radial Feature Dimensionality Reduction}
\label{sec:Radial}



It is non-trivial to determine the dimensionality $k$ of the radial feature vector $\mathbf g^{\mathsf{radial}}_{ij}$, because MD accuracy and accessible spatial scale must be carefully balanced. The former generally favors a larger $k$ to retain more information, whereas the latter favors a smaller $k$ to reduce the per-atom memory footprint. Traditionally, $k$ is determined empirically. DP always uses $k=100$~\cite{DP1,DP_compress,DP_Ppopp}, whereas \TensorMD uses $k=64$~\cite{TensorMD}.

In this work, we observe that shifting atom-species discrimination from only the regression network ($\mathcal R$ in Equation~\ref{eq:theory_E}) to the radial encoding network ($\mathcal E$ in Equation~\ref{eq:radial}) exposes a substantially lower rank radial feature space. This allows $k$ to be reduced further while retaining accuracy.

 \subsubsection{Differentiating Atom Species.}

Atom species discrimination can be achieved by letting  the encoding network $\mathcal{E}$ and the  regression network $\mathcal{R}$ to be differentiated by atom species,  via $\chi_\mathcal E$ and $\chi_\mathcal R$ (in Equations~\ref{eq:radial},~\ref{eq:theory_E}) respectively.
For a $C$-component system,  each can choose to carry different atom species information. 

  \begin{itemize}[leftmargin=10pt]
 \item
$\chi_\mathcal E$ can be configured as one of the following three cases:
\begin{itemize}
 \item
$\chi_\mathcal E = \emptyset$. The encoding network $\mathcal{E}$ is universal across all atom species.
\item
$\chi_\mathcal E = \{\mathbf s_j\}$. Differentiated by neighboring atom $j$'s  atom species $\mathbf s_j$, $C$ different encoding networks are used.
\item
$\chi_\mathcal E = \{(\mathbf s_i, \mathbf s_j)\}$. Differentiated by the  specie pair $(\mathbf s_i, \mathbf s_j)$ of cental atom $i$ and its neighbor $j$, $C^2$ different encoding networks are used.
\end{itemize}
 \item
$\chi_\mathcal R$ can be configured as one of the following two cases:
\begin{itemize}
 \item
$\chi_\mathcal R = \emptyset$. The regression network $\mathcal{R}$ is universal across all atom species.
\item
$\chi_\mathcal R = \{\mathbf s_i\}$. Differentiated by central atom $i$'s  atom specie $\mathbf s_i$, $C$ different regression networks are used.
\end{itemize}

\end{itemize}

 Though    existing MLIPs have chosen their own strategy to assign the atom species information to the encoding and regression networks -- e.g., \TensorMD adopts the combination of $\chi_\mathcal E = \emptyset \land \chi_\mathcal R = \{\mathbf s_i\}$, and \DPSE uses $\chi_\mathcal E = \{\mathbf s_j\} \land \chi_\mathcal R = \{\mathbf s_i\}$ --  however, how different combinations affect the radial feature vector length 
 $k$ and, in turn, influence the spatial scale of the MD simulation remains unaddressed.

\subsubsection{Exposing a Low-Rank Radial Feature Space.}

The physical intuition is that radial dependence is a smooth function on the bounded one-dimensional interval $[0,r_c]$. A universal encoder must provide one shared radial basis that remains useful for all chemical pairs, whereas atom-dependent conditioning separates only the neighbor species. Pair conditioning decomposes this mixed problem into smoother ordered-pair responses. Once the relevant bond-length and coordination-shell variations are resolved, additional channels tend to become correlated. This motivates, but does not by itself prove, a smaller radial rank.

We test the hypothesis by stacking radial vectors sampled over distance and species combinations into a matrix and applying singular value decomposition. Figure~\ref{fig:svd_rank}(b) reports the spectra from independently trained $k=64$, $\ell_{max}=3$ models for the three-component alloy $\mathsf{AlCuMg}$ and the six-component high-entropy alloy \HEA. Under pair conditioning, the first 21 and 29 singular directions capture $1-10^{-8}$ of the total squared singular value, respectively, whereas the universal and atom-dependent encoders require at least 48 directions. The faster decay confirms that pair conditioning exposes a substantially lower-rank radial feature space.

 \subsubsection{Determining the Radial Feature Dimensionality}

SVD identifies candidate widths but does not replace retraining. We consider $k\in\{16,32,64\}$, which also maps naturally to GPU tile and warp widths. For a material, a $k=64$ model is first analyzed; candidate narrower models are then trained from scratch and accepted only after accuracy, throughput, and memory evaluation. Retraining typically takes only a few hours on one NVIDIA A100 GPU from scratch. This one-time cost is amortized over production MD, where the smaller per-atom footprint can enable substantially larger simulations across many GPUs. Based on our experiences across various systems, $k=32$ is a balanced choice.

\subsection{Angular Feature Dimensionality Reduction}
\label{sec:Angular}

The canonical angular feature vector in \TensorMD\cite{TensorMD} contains all distinct Cartesian monomials of the unit direction $\hat{\mathbf r}_{ij}=(\hat{x}_{ij},\hat{y}_{ij},\hat{z}_{ij})$ through maximum angular moment $\ell_{max}$. The cumulative dimension of this reducible Cartesian representation, up to $\ell_{max}$, is
\begin{equation}
  d
  =\sum_{\ell=0}^{\ell_{max}}\binom{\ell+2}{2}
  =\frac{(\ell_{max}+1)(\ell_{max}+2)(\ell_{max}+3)}{6}.
  \label{eq:angular_reducible_dim}
\end{equation}
For example, $\ell_{max}=1$ gives $[1,\hat{x}_{ij},\hat{y}_{ij},\hat{z}_{ij}]$ and $d=4$. Extending to $\ell_{max}=2$ appends the six quadratic monomials $(\hat{x}_{ij}^2,\hat{x}_{ij}\hat{y}_{ij}, \hat{x}_{ij}\hat{z}_{ij},\hat{y}_{ij}^2, \hat{y}_{ij}\hat{z}_{ij},\hat{z}_{ij}^2)$ and gives $d=10$.


This basis is simple to generate but reducible under rotations. For example, the six quadratic monomials contain five independent quadrupolar components and the scalar trace $\hat x_{ij}^2+\hat y_{ij}^2+\hat z_{ij}^2=1$, which repeats information already present at $\ell=0$.


\subsubsection{Irreducible Harmonic Projection}

We remove such dependencies by applying the irreducible Harmonic projections on angular feature vector. 

 As in Equation~\ref{eq:angular}, $\mathbf g^{\mathsf{angular}}_{ij}$ is formed by concatenated  $\ell_{max}+1$ vectors $\{\mathbf g^{\ell}_{ij}|0\le \ell\le \ell_{max}\}$ together, with each $\mathbf g^{\ell}_{ij}$ containing all distinct Cartesian monomials of the unit direction
 $\hat{\mathbf r}_{ij}
 =(\hat{x}_{ij},\hat{y}_{ij},\hat{z}_{ij})$ whose total degree equals to $\ell$.

 An irreducible Harmonic projection is applied on each $\mathbf g^{\ell}_{ij}$ by left-multiplying it with a  matrix $Q_{\ell}\in\mathbb{R}^{(2\ell+1)\times \binom{\ell+2}{2}}$  satisfying $Q_{\ell}Q_{\ell}^{\mathsf T}=I$. The projection matrix $Q_{\ell}$ 
 is determined through a two-step procedure: first, redundant radial components are removed; second, the retained basis functions are enforced to be orthonormal on the sphere. Equivalently, $Q_{\ell}$ is obtained by solving the following two constraints:
\[
L_\ell Q_\ell^{\mathsf T}=0,
\qquad
Q_\ell D_\ell^{-1}Q_\ell^{\mathsf T}=I.
\]
where $L_{\ell}$ is the matrix representation of the Laplacian operator $\nabla^2$ mapping from degree $\ell$ homogeneous polynomials to degree $\ell-2$ polynomials, and $D_{\ell}\in \mathbb R^{\ell\times \ell}$ is a diagonal matrix, whose entries are the multinomial coefficients $\frac{\ell!}{\alpha_x!\,\alpha_y!\,\alpha_z!}$, with $\alpha_x,\alpha_y,\alpha_z$ the degrees of $\hat{x}_{ij},\hat{y}_{ij},\hat{z}_{ij}$ in $\mathbf g^{\ell}_{ij}$ respectively (thus satisfying $\alpha_x+\alpha_y+\alpha_z=\ell$).

\begin{table}[t]
\centering
\setlength{\tabcolsep}{4pt}
\small
\caption{Angular feature vector dimensions before and after irreducible reduction, up to $\ell_{max}$.}
\label{tbl:angular_feature}
\begin{tabular}{c|cccccc}
\toprule
\(\ell_{max}\) & 0 & 1 & 2 & 3 & 4 & 5 \\
\midrule
Original dimension & 1 & 4 & 10 & 20 & 35 & 56 \\
Reduced dimension  & 1 & 4 & 9 & 16 & 25 & 36 \\
Reduction rate & 0\% & 0\% & 10.0\% & 20.0\% & 28.6\% & 35.7\% \\
\bottomrule
\end{tabular}
\end{table}

\subsubsection{Choosing the Maximum Angular Order}
Similar to the radial feature dimensionality $k$, the maximum angular order $\ell_{max}$ -- thereby the angular feature dimensionality $d$ -- is determined empirically in existing MLIPs.

In practice, we have observed that $\ell_{max}=3$ usually provides a favorable balance between accuracy and efficiency across various materials, and this observation has a reasonable physical basis, with $\ell_{max}$ corresponding to the momentum quantum numbers in physical theory.
When $\ell_{max}=3$,  isotropic, dipolar, quadrupolar, and octupolar angular contents are encoded into $\ell=0,1,2,3$, respectively, and 
this angular coverage is sufficient for describing the structural diversity of bulk materials; including higher-order terms would bring only marginal gains.
Furthermore, as shown in Table~\ref{tbl:angular_feature}, $\ell_{max}=3$ 
corresponds to an angular feature dimensionality of 16, which is GPU-efficient.

\subsection{Intermediate Tensor Reduction}
\label{sec:Intermediate}

In this work, all intermediate tensors in the MLIP inference pipeline given in Figure~\ref{fig:fuse}(a), have been eliminated by mega-fusion, such that only a single mega-kernel is launched to the GPU per time-step of the MD simulation. Within such mega-kernel, all intermediate tensors are either kept on-chip (in registers or shared memory) or computed on-the-fly, thereby leaving no footprint on HBM.

This has been made possible by the dimensionality reduction introduced in Sections~\ref{sec:Radial}~\ref{sec:Angular}, which — by lowering  $k,d$ -- has significantly decreased the memory demand of the intermediate tensors.

\subsubsection{Configurable Fusion}

Table~\ref{tab:execution_strategies} shows two fusion configurations  available, for the purpose of balancing the MD simulation spatial scale and time-to-solution across GPUs with various on-chip memory capacities.

Before fusion, the original MLIP pipeline contains 10 CUDA kernels (corresponding to Equations~\ref{eq:setup}-\ref{eq:force}). Within each kernel, a single thread block processes a batch of atoms, with their vectors/tensors grouped together. 

To enable fusion, tiling -- reducing the batch of atoms processed by each thread block -- is necessary to keep the working set within the on-chip memory capacity. {During tiling, register pressure may  increase sharply,}
both mega-fusion and partial-fusion are thereby supported to avoid this.

\begin{itemize}[leftmargin=10pt]
\item \texttt{Mega-fusion}. Originally, the inference pipeline produces a large HBM footprint because the resultant vectors/tensors of each equation Equations 1–10) --i.e., $\mathbf g^{\mathsf{radial}}_{ij}$, $\mathbf g^{\mathsf{angular}}_{ij}$, $\mathbf P_i,\mathbf G_i, \mathbf W_i, \mathbf U_i, \mathbf V_i$ -- are required to be stored.
With mega-fusion, all  10 CUDA kernels are fused into a single one, achieving  minimal HBM footprint for MLIPs with all intermediate tensors eliminated.
Within such mega-kernel, the large-sized and long-live-range $\mathbf g^{\mathsf{radial}}_{ij}$, $\mathbf g^{\mathsf{angular}}_{ij}$ are re-computed in the \texttt{Deviation} stage, rather than being kept on-chip throughout from the \texttt{Encoding} stage.

\item \texttt{Partial-fusion}. The original 10 CUDA kernels, are fused into 3 (Equations 1-5; 6; 7-10, respectively), each responsible for one stage of the \texttt{Encoding}-\texttt{Regression}-\texttt{Deviation} pipeline.
Compared with \texttt{Mega-fusion}, results from Equations 5,6 are required to be stored on HBM, i.e., the descriptor $\mathbf G_i$ and the potential energy $E_i$ (committed in Table~\ref{tab:execution_strategies} since it is a scalar for each atom).
In addition to $\mathbf g^{\mathsf{radial}}_{ij}$ and $\mathbf g^{\mathsf{angular}}_{ij}$ as in \texttt{Mega-fusion}, $\mathbf P_i$ is also re-computed in  the \texttt{Deviation} stage.

\end{itemize}

The two fusion configurations offer flexibility to choose the appropriate one according to the GPU's on-chip memory capacity and the simulation precision. For example in this paper, the proposed MLIP is evaluated on NVIDIA A100 GPU, on which mega-fusion is adopted for FP32 simulations, whereas partial-fusion is used for FP64. This is because the lower memory footprint of FP32 allows a larger tiling size that fits comfortably within the register capacity, making mega-fusion particularly effective.

\begin{table}[t]
  \centering
  \small
  \setlength{\tabcolsep}{4.0pt}
  \caption{Configurable fusion  to eliminate  intermediate data.}
  \label{tab:execution_strategies}
  \begin{tabular}{@{}c|ccc@{}}
    \toprule
     {Fusion} & \# of CUDA & Sources of  & {Computed} \\
    configuration & kernels & HBM footprint & on-the-fly\\
    \midrule
     \multirow{2}{*}{\texttt{Mega-fusion}}    & 1  & \multirow{2}{*}{None}  & $\mathbf g^{radial}_{ij},$     \\
      & {\color{gray}(Eq 1-10, 1 total)} & &  $\mathbf g^{angular}_{ij}$\\
    \midrule
     \multirow{2}{*}{\texttt{Partial-fusion}} & 3 & \multirow{2}{*}{$\mathbf G_i$}   & $\mathbf g^{radial}_{ij}, $ \\
      & {\color{gray}(Eq 1-5; 6; 7-10)} &  & { $\mathbf g^{angular}_{ij},\mathbf P_i$}\\
    \midrule
     \multirow{2}{*}{Before fusion}  & 10 & $\mathbf g^{radial}_{ij}, \mathbf g^{angular}_{ij}, $ &  \multirow{2}{*}{None}   \\
     & {\color{gray}(Eq 1-10, 1 each)}  & $\mathbf P_i,\mathbf G_i, \mathbf W_i, \mathbf U_i, \mathbf V_i$ & \\

    \bottomrule
  \end{tabular}
\end{table}

\subsubsection{Tensor-Core-Accelerated Fusion}

In both fusion configurations, tensor cores accelerate the contractions underlying $\mathbf U_i$ and $\mathbf V_i$ in the backward stage (Eq~7--8). For each central atom, neighboring pairs are processed in tiles, and the radial features and their radial derivatives are multiplied by $\mathbf W_i^{\mathsf T}$ using warp-level matrix-multiply--accumulate instructions. The first product yields $\mathbf V_i$, whereas the second is obtained by reassociating the $\mathbf U_i$-dependent radial-force contraction, so the wider $\mathbf U_i$ tensor is never explicitly formed. The resulting tile-local vectors remain in shared memory and are immediately consumed by the following force projection, avoiding HBM traffic for these backward intermediates. 



\subsection{Implementation}
\label{sec:impl}
{The proposed MLIP is implemented with 20K lines of C/C++ and 17K lines of CUDA codes, and open-sourced}~\footnote{Link is hidden for double-blind review.}.

As with existing MLIPs, it interfaces with LAMMPS as the MD simulator. Previously, both DP~\cite{DP_GB,DP_Ppopp,DP_TC} and TensorMD~\cite{TensorMD} interfaced with the CPU version of LAMMPS, despite being GPU-accelerated themselves, because the MLIP computations dominate the overall workload, leaving the LAMMPS portion as a minor overhead.
However,  innovations above have significantly enhanced the performance of the proposed MLIP, thereby we further enable LAMMPS to use GPU, with its Kokkos interface.


\section{Evaluation}
\label{sec:eval}

\subsection{Experimental Setup}
\label{sec:setup}

\subsubsection{Platform Configuration}
\label{subsec:hardware}

A commercial GPU cluster of 144 NVIDIA A100 PCIe GPUs is used in evaluation. Each compute node contained two 64-core Armv8.2-A processors operating at 3.0~GHz and four NVIDIA A100 PCIe GPUs, each with 40~GB of HBM2 memory. Each node was connected to the cluster fabric through four 100-Gb/s RoCE links. The toolchains are: GCC 11.3.0, CUDA 12.2, CUDA-aware Open MPI 4.1.5 with UCX 1.15.0.

\subsubsection{Bulk-Material Systems}
\label{subsec:datasets}

We evaluated three bulk-material systems drawn from published studies: a unary metal, a ternary alloy, and a six-component high-entropy alloy (HEA), as summarized in Table~\ref{tbl:datasets}.

\begin{table}[t]
\centering
\small
\caption{Bulk-material systems used in the evaluation.}
\label{tbl:datasets}
\begin{tabular}{@{}lp{0.72\columnwidth}@{}}
\toprule
System & Description \\
\midrule
$\mathsf{Sn}$ &
Unary metallic system~\cite{DP_Sn} \\
\addlinespace
$\mathsf{AlCuMg}$ &
Ternary alloy~\cite{DP_compress,DPA1} \\
\addlinespace
$\mathsf{HEA}$ &
Six-component high-entropy alloy comprising
$\mathsf{Ta}$, $\mathsf{Nb}$, $\mathsf{W}$, $\mathsf{Mo}$,
$\mathsf{V}$, and $\mathsf{Al}$~\cite{DPA1} \\
\bottomrule
\end{tabular}
\end{table}

\begin{figure*}[t]
  \centering
  \small
  \begin{subfigure}[t]{0.51\textwidth}\vspace{0pt}
    \centering
    \small
    \begin{threeparttable}
      \setlength{\tabcolsep}{2pt}
      \begin{tabular}{c|l|cc|c|c|c}
      \toprule
       System & MLIP            & \multicolumn{2}{c|}{RMSE}      & Throughput       & Footprint      & \#atoms      \\
              &                 & E  & F   & (Matoms-step/s)   &   (KB/atom)   & ($\times 10^6$)  \\
      \midrule
       \multirow{3}{*}{\Sn}
        & \DPSE\cite{DP_Sn}     &     7.55 & 93.5  &  1.41            &     38.1      &  1.01          \\
        & \TensorMD             &     6.21          & 56.2           &  3.56            &     69.5      &  0.55          \\
        & \System               &  \textbf{4.69}    & \textbf{54.3}  & \textbf{15.75}   & \textbf{3.7}  & \textbf{10.98} \\
      \midrule
       \multirow{3}{*}{\AlCuMg}
        & \DPSE\cite{DPA1}      &    3.67  & 45    & 0.87             &      107.2    &  0.36          \\
        & \TensorMD             &    7.32           & 51.5           & 1.48             &     242.2     &  0.14          \\
        & \System               &  \textbf{3.20}    & \textbf{35.6}  & \textbf{9.33}    & \textbf{5.3}  &  \textbf{7.75} \\
      \midrule
       \multirow{3}{*}{\HEA}
        & \DPSE\cite{DPA1}      &    28.7  & 141  & 0.66             &     204.6     &  0.18          \\
        & \TensorMD             &   11.40           & 97.7           & 0.94             &     365.6     &  0.10          \\
        & \System               &    \textbf{2.75}  & \textbf{49.8}  & \textbf{9.15}    & \textbf{5.1}  &  \textbf{7.89} \\
      \bottomrule
      \end{tabular}
    \end{threeparttable}
    \caption{End-to-end MD simulation accuracy, throughput and spatial scale. The energy RMSE is reported in meV/atom, and the force RMSE in meV/\AA{}.}
    \label{tbl:compare}
  \end{subfigure}\hfill
  \begin{subfigure}[t]{0.47\textwidth}\vspace{0pt}
    \centering
    \includegraphics[width=\linewidth]{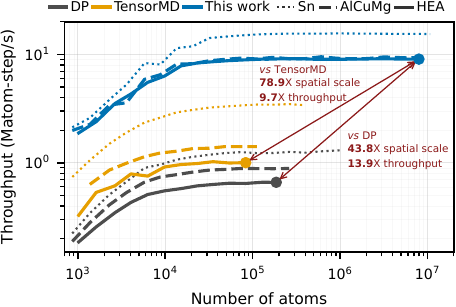}
    \caption{End-to-end MD  throughput and spatial scale until OOM.}
    \label{fig:saturation}
  \end{subfigure}


  \begin{subfigure}[t]{\textwidth}
    \centering
    \includegraphics[width=\linewidth]{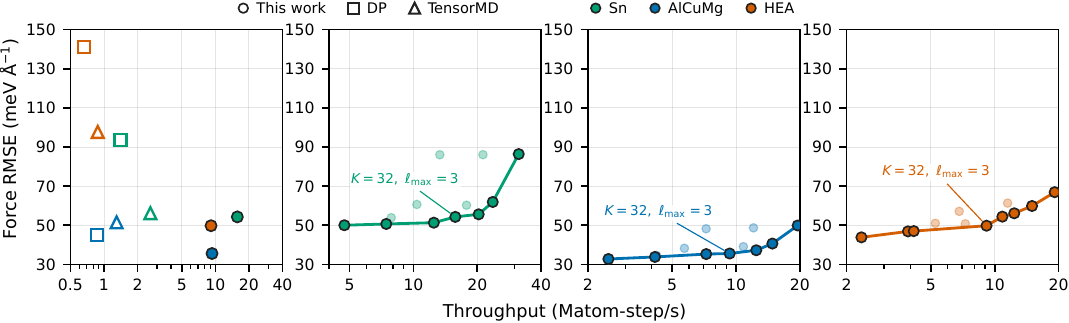}
    \caption{End-to-end MD accuracy-throughput  Pareto frontiers as the justification of defaulting $k=32,\ell_{max}=3$ in the proposed MLIP.}
    \label{fig:pareto_frontier}
  \end{subfigure}

  \caption{End-to-End MD simulation results on a single GPU of the proposed MLIP and the two baseline MLIPs, i.e., DP and \TensorMD{}, across three physical systems of the unary \Sn{}, the 3-component \AlCuMg{}, and the 6-component \HEA{}.}
  \label{fig:compare}
\end{figure*}

\subsubsection{MLIPs and Architectures}

For the proposed MLIP,  a pair conditioned shared encoder with three hidden layers of width 128, and an atom-species dependent shared energy head with four hidden layers of width 64, SiLU activations, and residual connections are used.

Two MLIPs were selected as the baselines for comparison.
\DPSE is evaluated using DeePMD-kit v3.1.2~\cite{DP_compress,DPkit-v3}, an open-source implemention with TensorFlow, for which model configurations  adopted from previous  studies~\cite{DP_Sn,DPA1} for all three physical systems. 
\TensorMD is a framework-independent MLIP, and  its published artifact~\cite{TensorMD_SC25_Artifact} is used under its default settings,  the same as in its publication~\cite{TensorMD}.

The cutoff radius is $r_c=6$~\AA{} for all three MLIPs except \DPSE/\Sn ($r_c=8$\AA{}\cite{DP_Sn}) to align with its published work. Unless otherwise specified, all calculations were performed in FP64.

\subsubsection{MD Simulation Configuration}

All implementations were coupled to LAMMPS through their respective pair styles. Simulations use a 1~fs timestep, and an NVT ensemble at 300~K. Neighbor lists were constructed using the binning algorithm with a 2~\AA{} skin. LAMMPS checked the neighbor-list rebuild criterion every five steps and rebuilt the list only when required.

\subsection{End-to-End Results on A Single GPU}

Figure~\ref{fig:compare}(a) summarizes the single-GPU comparison. Each row is measured at the maximum workable system size for that MLIP on a single NVIDIA A100 GPU, which is reported in the \#atoms column.
Results show that, the proposed MLIP has much smaller footprint than other MLIPs, consuming only 3.7--5.3~KB/atom, compared with 38.1--204.6~KB/atom for \DPSE and 69.5--365.6~KB/atom for \TensorMD. As a result, MD simulation spatial scale reaches 7.75--10.98 million atoms, 
whereas baselines are limited no more than 1 million atoms. 
Furthermore, performance benefits also come along, since the dimension reduction on feature vectors introduced in Section~\ref{sec:Radial} and ~\ref{sec:Angular} leads to computation strength reduction. 
The end-to-end MD throughput with the proposed MLIP reaches 9.15--15.75~Matom-step/s,
whereas baselines are lower than 4~Matom-step/s.  
On average, the proposed MLIP has improved throughput by 11.9$\times$ and 6.8$\times$, compared with \DPSE and \TensorMD respectively.
On accuracy, the proposed MLIP achieves smaller RMSE compared with \DPSE (results  taken from \cite{DP_Sn}, \cite{TensorMD}, \cite{DPA1}) and \TensorMD.

Figure~\ref{fig:compare}(b) reports the
 end-to-end MD throughput and spatial scale.
Results show that, the proposed MLIP can constantly achieve significantly higher throughput than baselines.
Take the 6-component \HEA for example, it can improve spatial scale by $43.8\times$ and  throughput by $13.9\times$ compared with \DPSE, and such improvements are $78.9\times$ and   $9.7\times$ respectively when compared with \TensorMD.

Figure~\ref{fig:compare}(c) reports the end-to-end MD  accuracy and throughput across  three MLIPs in the left sub-figure, with the six hollow \DPSE and \TensorMD{} markers lie above and to the left of the three solid markers of this work under the default $k=32,\ell_{max}=3$, indicating this work has advantages in both accuracy and throughput. 
Furthermore, the right three sub-figures give justification of the default configuration in this work, 
with 12 combinations of $k\in\{16,32,64\}$ and $\ell_{\max}\in\{1,2,3,5\}$ evaluated. Results show that $k=32,\ell_{max}=3$ lies on the Pareto frontiers (solid line) for all three systems, with excellent balance between accuracy and throughput.

\begin{table}[t]
\small
\centering
\setlength{\tabcolsep}{4pt}
\caption{End-to-end MD simulation results under  radial feature  dimensionalities $k=16,32,64$. The energy RMSE is reported in meV/atom, and the force RMSE in meV/\AA{}.}
\label{tbl:radial_ablation}
\begin{tabular}{c|c|cc|c|c}
\toprule
    \multirow{2}{*}{System}    & \multirow{2}{*}{$k$} & \multicolumn{2}{c|}{RMSE} & Throughput     & Footprint  \\
              &     &       E   &    F          & (Matom-step/s) & (KB/atom) \\
\midrule
\multirow{3}{*}{\Sn}
& 16 & 4.83 & 55.6 & 20.29 & 3.1 \\
& 32 & 4.69 & 54.3 & 15.75 & 3.7 \\
& 64 & 4.68 & 53.9 & 7.85  & 4.8 \\
\midrule
\multirow{3}{*}{\AlCuMg}
& 16 & 3.59 & 37.3 & 12.46  & 4.7 \\
& 32 & 3.20 & 35.6 & 9.33   & 5.3 \\
& 64 & 2.95 & 34.6 & 4.22   & 6.3 \\
\midrule
\multirow{3}{*}{\HEA}
& 16 & 3.31 & 56.1 & 12.36  & 4.6 \\
& 32 & 2.75 & 49.8 & 9.15   & 5.1 \\
& 64 & 2.77 & 46.9 & 3.90   & 6.2 \\
\bottomrule
\end{tabular}
\end{table}

\begin{table}[t]
\small
\centering
\setlength{\tabcolsep}{4pt}
\caption{End-to-end simulation results before and after angular feature dimensionality reduction under $k=32,\ell_{max}=3$. }
\label{tbl:angular_ablation}
\begin{tabular}{c|c|cc|c}
\toprule
\multirow{2}{*}{System} & \multirow{2}{*}{$d$}
  & \multicolumn{2}{c|}{RMSE}
  & Throughput \\
& & E (meV/atom) & F (meV/\AA{}) & (Matom-step/s) \\
\midrule
\multirow{2}{*}{\Sn}
& 20 & 4.91 & 55.1 & 8.86   \\
& 16 & 4.69 & 54.3 & 15.75  \\
\midrule
\multirow{2}{*}{\AlCuMg}
& 20 & 3.18 & 36.0 & 4.97  \\
& 16 & 3.20 & 35.6 & 9.33  \\
\midrule
\multirow{2}{*}{\HEA}
& 20 & 2.86 & 52.2 & 4.89  \\
& 16 & 2.75 & 49.8 & 9.15  \\
\bottomrule
\end{tabular}
\end{table}

\begin{figure}[t]
  \centering
  \includegraphics[width=0.9\columnwidth]{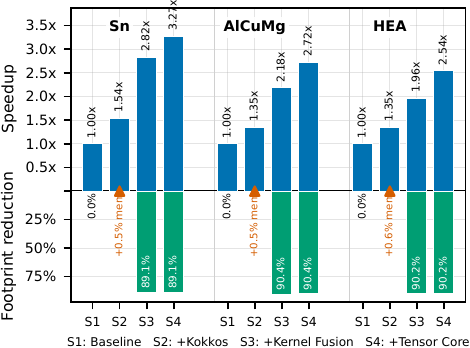}
  \caption{Performance improvement and footprint reduction achieved by eliminating intermediate tensors.}
  \label{fig:step-by-step}
\end{figure}

\subsection{Ablation Studies}


\subsubsection{ Radial Feature Dimensionality Reduction}
Table~\ref{tbl:radial_ablation} shows the result of radial  dimensionality reduction  (Section~\ref{sec:Radial}) under  $\ell_{max}=3$. 
Results show that  reducing the radial dimensionality $k$ from 64 to 32 raises throughput by 2.00--2.35$\times$ and reduces footprint by 0.9-1.0KB/atom, while force RMSE changes by only 0.8--6.2\%.  
Reducing $k$ from 32 to 16 increases force RMSE by a further 2.4--12.8\%, but adds only 1.29--1.35$\times$ throughput and reduces footprint  by only 0.5-0.6KB/atom. 
Thereby, we have identified that $k=32$ gives the most consistent balance across all three systems.

\subsubsection{   Angular Feature Dimensionality Reduction}
\label{sec:ablation-angular}

Table~\ref{tbl:angular_ablation} shows the result of angular  dimensionality reduction  (Section~\ref{sec:Angular}), under    $k=32, \ell_{max}=3$. 
Before reduction, the angular  dimensionality is $d=20$, and after reduction, $d=16$. 
The mean force RMSE differences are small: within 1.4-3.3\% variance. 
Lower RMSE can be achieved with $d=16$ since removing linear dependencies allows the energy head to use its capacity more effectively. The primary benefit is systems performance, with end-to-end throughput improving by factors of 1.76$\times$--1.86$\times$.


\subsubsection{Intermediate Tensor Elimination}

Figure~\ref{fig:step-by-step} shows the results of eliminating intermediate tensors   (Section~\ref{sec:Inference_Pipeline}), when simulating 864,000 \Sn{} atoms, 500,000  \AlCuMg{} atoms, and 524,288 \HEA{} atoms, respectively. 
In this figure, the un-fused MLIP interfacing with CPU-only LAMMPS is used as  baseline. By gradually including GPU-accelerated LAMMPS, \texttt{partial-fusion} (for this FP64 evaluation), and tensor-core acceleration, 2.54-3.27$\times$ of speedups can be observed, with HBM footprint reduced by $89.1-90.4\%$.



\subsection{Scaling Results}




\subsubsection{Weak Scaling.}
\begin{figure*}[t]
  \centering
  \begin{subfigure}{\textwidth}
    \centering
    \includegraphics[width=\textwidth]{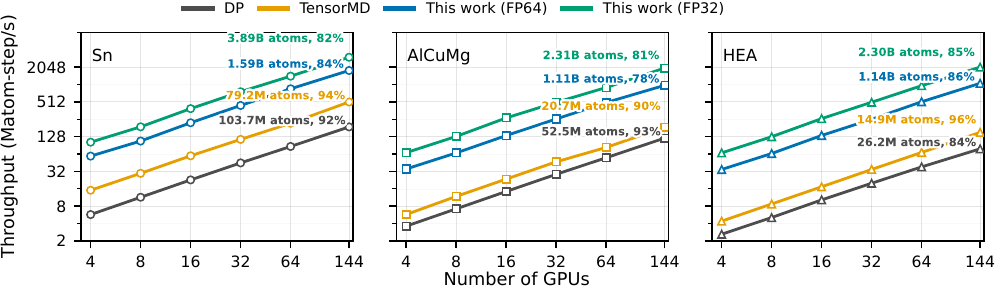}
    \caption{Weak scaling results with the maximum per‑GPU atom count (without OOM) across three systems.}
    \label{fig:weak_scaling}
  \end{subfigure}


  \begin{subfigure}{\textwidth}
    \centering
    \includegraphics[width=0.95\textwidth]{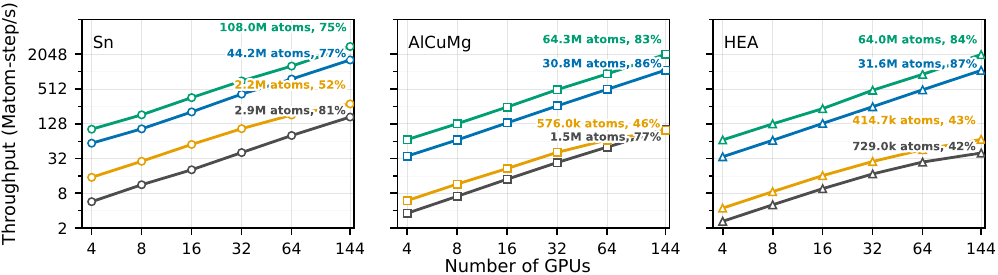}
    \caption{Strong scaling results with atom counts fixed to $4\times$ of the per-GPU counts used in \textbf{(a)}, to match the minimum 4‑GPU configuration.}
    \label{fig:strong_scaling}
  \end{subfigure}
  \caption{Weak and Strong scaling results from 4 to 144 GPUs across three physical systems of the unary \Sn{}, the 3-component \AlCuMg{}, and the 6-component \HEA{}, with three MLIPs of \DPSE, \TensorMD and this work. Results are reported as absolute end-to-end MD throughput in Matom-step/s, and parallel efficiencies and MD spatial scales on 144 GPUs are explicitly marked.}
  \label{fig:scaling}
\end{figure*}

Figure~\ref{fig:scaling}(a) shows the weak-scaling results from 4  to 144 GPUs, with the maximum per‑GPU atom count (without OOM) across \Sn{}, \AlCuMg{} and \HEA{}. 

For the unary \Sn{} system, the processed atoms on each GPU are 550,000 and 11,054,400 in \TensorMD and this work, respectively, the same as the maximum  counts reported in Figure~\ref{fig:compare}(a). While  \DPSE{} uses 720,000  atoms/GPU, below the single‑GPU maximum shown in Figure~\ref{fig:compare}(a), because scaling to the 144‑GPU cluster triggers OOM beyond this size.
Similarly configured, in the 3-component \AlCuMg{} simulation, 364,500/144,000/7,688,000 atoms are processed on each GPU with 
\DPSE{}/\TensorMD{}/this work respectively;
and in the 6-component \HEA{} simulation, the per-GPU-processed-atoms are 182,250/103,680/7,888,624.

Results show that, the proposed MLIP can achieve significant larger MD spatial scale and higher simulation throughput, under  satisfying weak-scaling efficiencies.
Take the 6-component \HEA{} for example, $1.14\times 10^9$ atoms can be simulated on 144 GPUs, while the two baseline MLIPs can only process $1.49\times 10^7$ and $2.62\times 10^7$ atoms. Under these configurations, the MD simulation throughput achieves 1072.65 Matoms-step/s for this work, $13.62\times$ and $7.06\times$ of \DPSE and \TensorMD respectively.
While the  weak-scaling efficiencies,  $86\%$/$96\%$/$84\%$ for this work/\TensorMD/\DPSE,  are all satisfying.

\begin{table}[t]
\centering
\small
\setlength{\tabcolsep}{3pt}
\renewcommand{\arraystretch}{0.95}
\caption{MPI task  breakdown of this work on the \HEA{} system, for the  weak scaling results reported in Figure~\ref{fig:scaling}(a), at 4, 32, 144~GPUs.  ``Comm'', ``Compute'' and ``Other'' represent the communications, computations and others in LAMMPS.}
\label{tab:mpi-scaling}
\resizebox{\columnwidth}{!}{%
\begin{tabular}{l cc cc cc}
\toprule
 & \multicolumn{2}{c}{4~GPUs} & \multicolumn{2}{c}{32~GPUs} & \multicolumn{2}{c}{144~GPUs} \\
\cmidrule(lr){2-3} \cmidrule(lr){4-5} \cmidrule(lr){6-7}
Section & Time (s) & \% & Time (s) & \% & Time (s) & \% \\
\midrule
MLIP   & 84.10 & 92.13 & 84.98  & 86.15 & 84.92  & 80.18 \\
Comm   & 6.24  & 6.84  & 10.62  & 10.77 & 16.31  & 15.40 \\
Compute & 0.12  & 0.13  & 2.09   & 2.12  & 3.55   & 3.35  \\
Other  & 0.83  & 0.91  & 0.94   & 0.95  & 1.13   & 1.06  \\
\midrule
Total  & 91.28 & 100.00 & 98.64  & 100.00 & 105.90 & 100.00 \\
\bottomrule
\end{tabular}%
}
\end{table}

Table~\ref{tab:mpi-scaling} shows the end-to-end MD simulation time breakdown of the proposed MLIP, when simulating \HEA on 4, 32, 144 GPUs respectively. Results show that, as the number of GPUs grows, the time share of the proposed MLIP drop from $92.13\%$ to $80.18\%$, with more time spent in LAMMPS. On one hand, all MPI communications are handled by LAMMPS, on the other, computations in LAMMPS are responsible to update the atomic coordinates in each step, both increase as the numbers of GPUs and atom counts grow.

Besides FP64 evaluations above, FP32 is also evaluated for this work, which can approximately double the MD spatial scale and  throughput, with similar parallel efficiencies.


\subsection{Strong Scaling.}
Figure~\ref{fig:scaling}(b) shows the strong scaling results from 4 to 144 GPUs, with the atom counts fixed to $4\times$ of the per-GPU counts used in the weak-scaling evaluation, to match the minimum 4-GPU configuration.

In the evaluation,
\DPSE{} simulates 2,880,000 \Sn{}, 1,458,000 \AlCuMg{}, and 729,000 \HEA{} atoms; \TensorMD{} simulates 2,200,000 \Sn{}, 576,000 \AlCuMg{}, and 414,720 \HEA{} atoms; and \system simulates 44,217,600 \Sn{}, 30,752,000 \AlCuMg{}, and 31,554,496 \HEA{} atoms.
At 144 GPUs, \system retains $76.8\%$, $85.6\%$, and $86.9\%$ parallel efficiency for \Sn{}, \AlCuMg{}, and \HEA{}, respectively, outperforming \TensorMD{} on all three systems ($51.9\%$, $46.0\%$, and $42.9\%$). \System also outperforms DP on \AlCuMg{} and \HEA{} ($77.1\%$ and $42.3\%$), while \Sn{} is the exception, where DP retains $81.2\%$.

From Figure~\ref{fig:compare}(b), we can discover the reason of excellent strong scaling of this work compared with the two baselines. The proposed MLIP can retain a high MD  throughput  when simulating a small number of atoms, whereas the two baselines only achieves peak  throughput when the system size approaches their maximum workable size.


Similar to weak scaling, the proposed MLIP can double the MD throughput when using FP32,  compared to FP64 above.

\section{Conclusion}
\label{sec:conclusion}

{In this paper, we have proposed an MLIP with a significantly reduced per-atom memory footprint, which enables MD simulations at the $\sim 10^9$ atom scale to be completed using only 144 GPUs, making practical bulk material research -- which previously required tens of thousands of GPUs -- feasible on commonly accessible clusters. This has been achieved by innovations that address the the two primary contributors to HBM footprint, i.e., feature vectors and intermediate variables, respectively.

\bibliographystyle{ACM-Reference-Format}
\bibliography{ref}

\end{document}